\documentclass[aps,reprint,superscriptaddress,citeautoscript,floatfix,longbibliography,bibnotes]{revtex4-2}
\usepackage{physics,times,bm,braket,graphicx,amsmath,amsfonts,amssymb,hyperref}
\usepackage[utf8]{inputenc}
\usepackage[T1]{fontenc}
\hypersetup{colorlinks=true,urlcolor=blue,linkcolor=blue,citecolor=red}
\usepackage[capitalize]{cleveref}
\usepackage{float}
\newcommand{\ii}{{\mathrm i}}
\begin{document}
\title{
Controlling energy spectra and skin effect via boundary conditions in non-Hermitian lattices
}
	\author{S Rahul}
		\affiliation{Department of Science and Humanities, PES University EC Campus, Bangalore 560100, India}
\author{
Pasquale Marra
}
\email{pasquale.marra@keio.jp}
\affiliation{
Department of Engineering and Applied Sciences, Sophia University, 7-1 Kioi-cho, Chiyoda-ku, Tokyo 102-8554, Japan
}
\affiliation{
Department of Physics \& Research and Education Center for Natural Sciences, Keio University, 4-1-1 Hiyoshi, Yokohama, Kanagawa, 223-8521, Japan
}
\affiliation{
Graduate School of Informatics, Nagoya University, Furo-cho, Chikusa-Ku, Nagoya, 464-8601, Japan
}
	\date{\today}
	
\begin{abstract}
Non-Hermitian systems exhibit unique spectral properties, including the non-Hermitian skin effect and exceptional points, often influenced by boundary conditions.
The modulation of these phenomena by generalized boundary conditions remains unexplored and not understood. 
Here, we analyze the Hatano-Nelson model with generalized boundary conditions induced by complex hopping amplitudes at the boundary. 
Using similarity transformations, we determine the conditions yielding real energy spectra and skin effect, and identify the emergence of exceptional points where spectra transition from real to complex. 
We demonstrate that tuning the boundary hopping amplitudes precisely controls the non-Hermitian skin effect, i.e., the localization of eigenmodes at the lattice edges. 
These findings reveal the sensitivity of spectral and localization properties to boundary conditions, providing a framework for engineering quantum lattice models with tailored spectral and localization features, with potential applications in quantum devices.
\end{abstract}
\maketitle


Non-Hermitian systems have attracted significant attention due to their rich spectral properties and the emergence of phenomena such as the non-Hermitian skin effect and exceptional points~\cite{bender_pt-symmetric_1999,heiss_the-physics_2012,bender_pt-symmetric_2015,yuto-ashida_non-Hermitian_2020,kawabata_topological_2021,okuma_non-Hermitian_2023}. 
In these systems, the presence of $\mathcal{PT}$ symmetry~\cite{bender_pt-symmetric_1999,mostafazadeh_pseudo-hermiticity_2002,bender_pt-symmetric_2015,PhysRevLett.80.5243, Bender_2007}, i.e., the simultaneous combination of spatial inversion (parity) and time-reversal symmetry, enforces the presence of real spectra with exceptional points marking the transition between phases with unbroken and broken $\mathcal{PT}$ symmetry.
These exceptional points are degeneracies in parameter space where both eigenvalues and eigenvectors coalesce~\cite{heiss_the-physics_2012,RevModPhys.93.015005}. 
A striking feature of these phases is the presence of the non-Hermitian skin effect, where bulk eigenmodes become localized at one boundary of a system with open boundary conditions, in contrast to the extended bulk states of Hermitian systems.
These systens also show a crossover between extended modes and skin modes when continuously interpolating between open and closed boundary conditions~\cite{edvardsson_sensitivity_2022}.
The non-Hermitian skin effect can be described as a direct consequence of nontrivial topology~\cite{okuma_topological_2020,zhang_correspondence_2020,davies_two-scale_2025, Gohsrich_2025} or, as recently proposed, as a manifestation of an effective spacetime curvature~\cite{marra_metric-induced_2024,marra_nonhermitian_2026}.

Among non-Hermitian systems, the Hatano-Nelson model~\cite{hatano_localization_1996,hatano_vortex_1997,hatano_localization_1998, okuma_non-Hermitian_2023, longhi_selective_2022, li_impurity_2021}, is a paradigmatic example, describing a one-dimensional lattice where non-Hermiticity arises from the asymmetry between hopping amplitudes in the left and right directions.
When the lattice is closed into a loop, enforcing periodic boundary conditions (PBC), this asymmetry leads to a complex-valued spectrum with eigenmodes characterized by well-defined momenta.
In contrast, for open boundary conditions (OBC), this asymmetry leads to real spectra but with eigenvalues exponentially localized at the two opposite edges, which is the manifestation of the non-Hermitian skin effect.
This behavior can be understood in terms of an imaginary gauge transformation that maps the non-Hermitian Hamiltonian to a Hermitian one~\cite{hatano_localization_1996,hatano_vortex_1997,hatano_localization_1998,okuma_non-Hermitian_2023}.

Here, we go beyond the usual PBC and OBC to focus on generalized boundary conditions (GBC), analyzing their impact on spectral properties such as the reality of the eigenvalues, the emergence of exceptional points, and the localization of the eigenmodes.
These GBC are implemented by introducing a weak or strong link between the two sites at the boundary of the lattice having an arbitrary complex phase, interpolating between periodic, anti-periodic, and open boundary conditions.
This model also describes a system with periodic boundary conditions but with a "defective" link localized between two lattice sites, describing a single impurity in an otherwise translational invariant system.
We find that both the reality of the eigenvalues and the localization of the eigenmodes are extremely sensitive to the hopping amplitudes and the complex phase describing the boundary link.
Hence, this enable a precise control of non-Hermitian phenomena, where the reality of the spectrum and the localization of the skin modes can be controlled by tuning only one parameter, which is convenient for experimental realization.
This also reveals an unusual dependence of the spectral features on the system size, since the condition for reality turns out to be size-dependent.

\begin{figure}
	\includegraphics[width=1\linewidth]{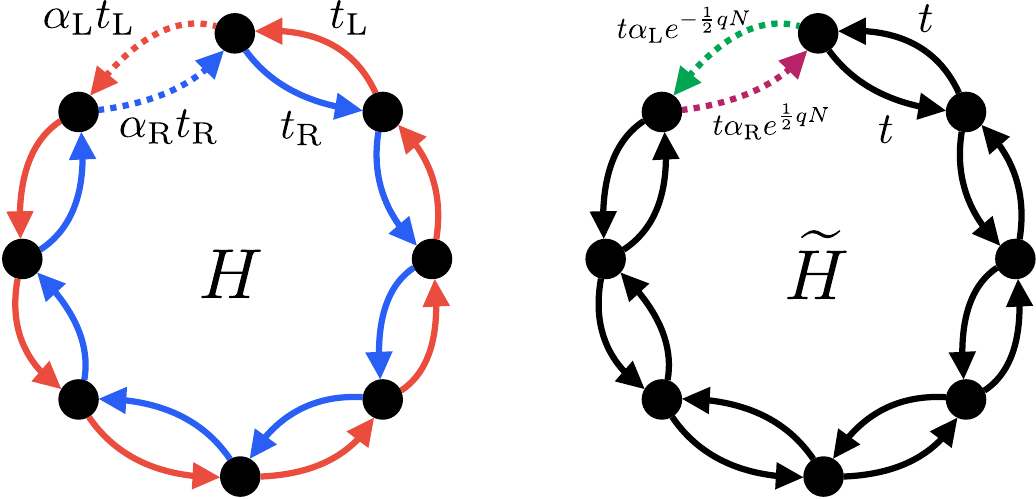}				
	\caption{
A sketch of the two isospectral Hamiltonians $H$ in \cref{eq:Hamiltonian} and $\widetilde H$ in \cref{eq:TransformedHamiltonian}.
The non-Hermitian Hamiltonian $H$ coincides with the Hatano-Nelson model with generalized boundary conditions or, alternatively, in the presence of a defective link.
The Hermitian Hamiltonian $\widetilde H$ coincides with that of a lattice fermion with generalized boundary conditions or in the presence of a defective link.
}
	\label{fig:chain}
\end{figure}

Specifically, we consider the Hatano-Nelson model without disorder and with generalized boundary conditions, described by the Hamiltonian
\begin{align}
H = \sum_{n=1}^{N-1}&
 \left(t_{\rm R} c^{\dag}_{n} c^{}_{n+1} + t_{\rm L} c^{\dag}_{n+1} c^{}_{n} \right)
\nonumber\\
+ &\left(\alpha_{\rm R} t_{\rm R} c^{\dag}_{N} c^{}_{1} + \alpha_{\rm L} t_{\rm L} c^{\dag}_{1} c^{}_{N} \right)
,
\label{eq:Hamiltonian}
\end{align}
$c^{\dag}_{n}$ and $ c^{}_{n}$ are creation and annihilation operators, $t_{\rm L},t_{\rm R}\in\mathbb{C}$ the left and right hopping amplitudes, and $\alpha_{\rm L}$ and $\alpha_{\rm R}$ the parameters that modulates the hopping amplitudes between the lattice sites $N$ and $1$, thereby controlling the boundary conditions at the two ends of the chain. 


\begin{figure}
	\includegraphics[width=1\linewidth]{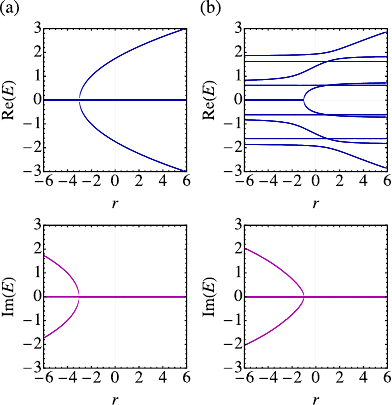}	
	\caption{
The energy spectra with exceptional points of the two isospectral Hamiltonians $H$ and $\widetilde H$ with
$\alpha_{\rm L}=r e^ {\frac{1}{2} q N }$
$\alpha_{\rm R}= e^ {-\frac{1}{2} q N }$ as a function of $r$ for lattices with $N=4$ (a) and $N=10$ (b), with $q=4$.
}
	\label{fig:EP}
\end{figure}

\begin{figure}
	\includegraphics[width=1\linewidth]{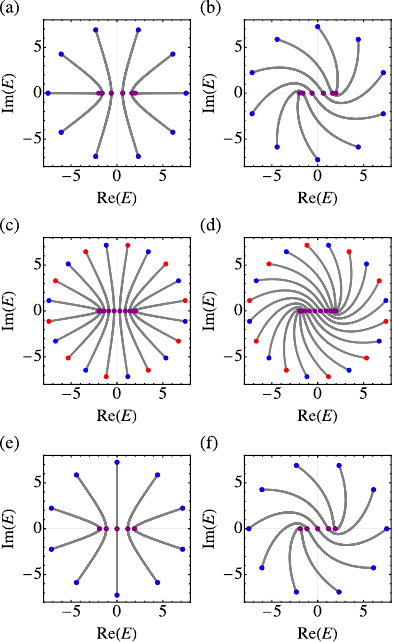}	
	\caption{
Energy spectra of the two isospectral Hamiltonians $H$ and $\widetilde H$ in \cref{eq:Hamiltonian,eq:TransformedHamiltonian}) on a lattice with $N=10$ lattice sites and with $\alpha_{\rm L}$ and $\alpha_{\rm R}$ as in \cref{eq:alphas} as a function of $\rho$ varying continuously in the interval $[0,q/2]$ and for different choices of $\phi$ and $q$.
We take 
$\phi=0$ [first row, (a), (b)],
$\phi=\pi/2$ [second row, (c), (d)],
$\phi=\pi$ [third row, (e), (f)],
with
$q=4$ [first column, (a), (c), (e)], 
$q=4+\ii\pi$ [second column, (b), (d), (f)].
Note that 4 out of $N=10$ eigenvalues are doubly degenerate. 
The highlighted points correspond to $\rho=0,1,2$, with the exceptional point $\rho=1$ corresponding to a purely real energy spectrum, in agreement with the condition in \cref{eq:specialBC}.
The trajectories followed by the complex eigenvalues as a function of $\rho$ swirl about the origin with a rotation angle that increases with $\Im(q)$, and their topology is affected by the choice of $\phi$.
}
	\label{fig:spectraeven}
\end{figure}

\begin{figure*}
	\includegraphics[width=1\linewidth]{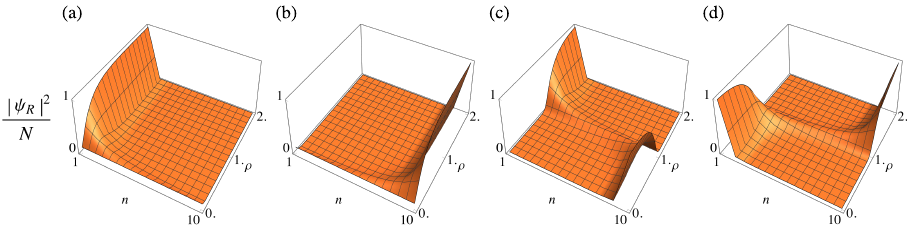}						
	\caption{
Average probability density for the right and left eigenmodes of the Hamiltonians $H$ and $\widetilde H$ in \cref{eq:Hamiltonian,eq:TransformedHamiltonian} on a lattice with $N=10$ lattice sites, with $\alpha_{\rm L}$ and $\alpha_{\rm R}$ as in \cref{eq:alphas} as a function of $\rho$ varying continuously in the interval $[0,2]$ with $q=4$ and $\phi=0$.
(a) Right eigenmode and (b) left eigenmode of $H$.
The eigenmodes of $H$ exhibit skin effect with exponential localization at the boundary for any $\rho\neq0$.
(c) Right eigenmode and (d) left eigenmode of $\widetilde H$.
In this case, the eigenmodes of $\widetilde H$ exhibit skin effect with exponential localization at the boundary for any $\rho\neq1$.
The right and left eigenmodes of the Hamiltonians $H$ and $\widetilde H$ are given by \cref{eq:eigenmodes}.
Different choices of the parameters $q$, $\phi$, and of the lattice size $N$ lead to qualitatively similar outcomes. 
}
	\label{fig:skin}
\end{figure*}

The Hamiltonian in \cref{eq:Hamiltonian} is invariant under $\mathcal{PT}$ symmetry~\cite{bender_pt-symmetric_1999,mostafazadeh_pseudo-hermiticity_2002,bender_pt-symmetric_2015}.
Indeed, if we define the space-inversion symmetry via the operator $\mathcal{P}$ acting on the lattice site indexes as $n\to N+1-n$ and the time inversion $\mathcal{T}$ as the antiunitary complex conjugation, one can directly verify that the Hamiltonian is invariant under the combined action of these two symmetries $\mathcal{PT}$, i.e., $\mathcal{H}=\mathcal{PT}\mathcal{H}(\mathcal{PT})^{-1}=\mathcal{P}\mathcal{H}^*\mathcal{P}$.
The key question is thus whether this $\mathcal{PT}$-symmetry is unbroken (leading to real energy spectra) or spontaneously broken (leading to complex spectra).
By applying the similarity transformation~\cite{hatano_localization_1996,hatano_vortex_1997,hatano_localization_1998,okuma_non-Hermitian_2023} defined by
\begin{equation}
 c^{\dag}_{n} \to 
 e^{\frac12qn} \tilde{c}^{\dag}_{n}
 ,
\quad
 c^{}_{n} \to 
 e^{-\frac12qn} \tilde{c}^{}_{n},
\end{equation}
where ${t_{\rm R}}/{t_{\rm L}}=e^{q}$ with $q\in\mathbb{C}$, yields
\begin{align}
\widetilde{H} = 
&
\sum_{n=1}^{N-1} t\left( \tilde{c}^{\dag}_{n} \tilde{c}^{}_{n+1} + \tilde{c}^{\dag}_{n+1} \tilde{c}^{}_{n} \right) 
\nonumber\\
+
&
t
\left[
\alpha_{\rm R} 
e^{\frac12qN}
\tilde{c}^{\dag}_{N} \tilde{c}^{}_{1}
+
\alpha_{\rm L} 
e^{-\frac12qN}
\tilde{c}^{\dag}_{1} \tilde{c}^{}_{N} 
\right],
\label{eq:TransformedHamiltonian}
\end{align}
with $t=\sqrt{t_{\rm L} t_{\rm R}}$.
The Hamiltonian $\widetilde{H}$ is isospectral (but not unitarily equivalent) to the Hamiltonian $H$ in \cref{eq:Hamiltonian}.

Note that the boundary conditions on the two Hamiltonians $H$ and $\widetilde H$ do not coincide in general. 
The Hamiltonian $H$ has 
open boundary conditions (OBC) for $\alpha_{\rm L}=\alpha_{\rm R}=0$,
periodic boundary conditions (PBC) for $\alpha_{\rm L}=\alpha_{\rm R}=1$, and 
antiperiodic boundary conditions (APBC) for $\alpha_{\rm L}=\alpha_{\rm R}=-1$.
Other values of $\alpha_{\rm L}$ and $\alpha_{\rm R}$ correspond to generalized boundary conditions (GBC) in a loop geometry, for instance, with a "weak" or a "strong" link between the lattice sites $N$ and $1$ in the case $|\alpha_{\rm L,R}|<1$ or $|\alpha_{\rm L,R}|>1$, respectively.
Conversely, the Hamiltonian $\widetilde H$ has 
OBC for $\alpha_{\rm L}=\alpha_{\rm R}=0$,
PBC for $\alpha_{\rm L}=1/\alpha_{\rm R}=e^{\frac12qN}$, and
APBC for $\alpha_{\rm L}=1/\alpha_{\rm R}=-e^{\frac12qN}$, and GBC otherwise.
The two lattice Hamiltonians are sketched in \cref{fig:chain}.
Note that these models can also be interpreted as lattice Hamiltonians with periodic boundary conditions, with a "defective" link localized between the lattice sites $N$ and $1$.
This defective link can describe a single impurity in an otherwise translational invariant system.

When the Hamiltonian $H$ has OBC ($\alpha_{\rm L}=\alpha_{\rm R}=0$), then $\widetilde{H} = \sum_{n=1}^{N-1} t\left( c^{\dag}_{n} c^{}_{n+1} + c^{\dag}_{n+1} c^{}_{n} \right)$, which is a Hermitian Hamiltonian (thus having real energy spectra) describing lattice fermions with OBC.
Conversely, when the Hamiltonian $H$ has PBC ($\alpha_{\rm L}=\alpha_{\rm R}=1$) or APBC ($\alpha_{\rm L}=\alpha_{\rm R}=-1$), the transformed Hamiltonian $\widetilde{H}$ is not Hermitian, since the hopping amplitudes at the end of the chain (second line of \cref{eq:TransformedHamiltonian}) are not reciprocal (and nonzero).
In these cases, the Hamiltonian has complex energy spectra and can be diagonalized in terms of momentum eigenmodes as
$\widetilde{H} = \sum_k \left( t_{\rm L} e^{\ii k} + t_{\rm R} e^{-\ii k} \right) c^\dag_k c^{}_k$ with $k=2\pi n/N$ for PBC and $k=\pi (2n+1)/N$ for APBC.

However, the choice of OBC is not a necessary condition to obtain real spectra.
Indeed, imposing the condition
\begin{equation}\label{eq:specialBC}
\alpha_{\rm L}=1/\alpha_{\rm R}=
e^{\ii\phi}
e^{\frac12qN},
\end{equation}
with $\phi\in[0,2\pi)$,
the transformed Hamiltonian $\widetilde{H}$ in \cref{eq:TransformedHamiltonian} reduces to $\widetilde{H} = \sum_{n=1}^{N}t\left( c^{\dag}_{n} c^{}_{n+1} + c^{\dag}_{n+1} c^{}_{n} \right)$, by identifying $c^{}_{N+1}\equiv e^{\ii\phi} c^{}_{1}$, thus obtaining a Hermitian Hamiltonian $\widetilde H$ with real energy spectra.
In terms of the phase $\phi$,
the Hamiltonian $H$ has PBC for $\phi-\frac\ii2 q N\equiv0\mod{2\pi}$ and APBC for $\phi-\frac\ii2 qN \equiv\pi\mod{2\pi}$, assuming $\Re(q)=0$.
Conversely, the Hamiltonian $\widetilde H$ has PBC for $\phi=0$ and APBC for $\phi=\pi$.

The fact that the the spectrum is real when $\alpha_{\rm L}=\alpha_{\rm R}=0$ or when $\alpha_{\rm L}$ and $\alpha_{\rm R}$ satisfy \cref{eq:specialBC}, and become complex when $\alpha_{\rm L}=\alpha_{\rm R}=\pm1$, indicates the existence of exceptional points where the spectra transitions from real to complex.
For example, for $N=4$, the energy spectrum is 
$E=\pm\frac{t}{\sqrt 2}{\sqrt{
\alpha^2
+3 
\pm\sqrt{
\Delta}
}
}$, where 
$\Delta=\alpha^4+2 \alpha^2 +
4 ({\alpha_{\rm L} e^{-2 q}}+ { \alpha_{\rm R} e^{2 q}})
+5$ and $\alpha^2=\alpha_{\rm L} \alpha_{\rm R}$.
This leads to a square root singularity as a function of $\alpha_{\rm L}$ and $\alpha_{\rm R}$, with an exceptional point occurring at $\Delta=0$. 
\Cref{fig:EP} shows the energy spectrum and the presence of exceptional points as a function of the parameter $\alpha_{\rm L}$ for different lattice sizes.
Notably, this provides an example where the splitting of eigenvalues at the exceptional points is controlled by hopping amplitudes that only affect the boundary.

Furthermore, we find that for $\alpha_{\rm L},\alpha_{\rm R}\neq0$ and
\begin{equation}\label{eq:alphas}
\alpha_{\rm L}=1/\alpha_{\rm R}=e^{\ii\phi}e^{\frac12 \rho q N},
\end{equation}
with $\rho \in\mathbb{R}$, the right and left eigenmodes of the Hamiltonian $H$ and $\widetilde H$ are
\begin{subequations}\label{eq:eigenmodes}
\begin{align}
\ket{k}_{\rm R}=&
\sum_{n=1}^N e^{-\frac12qn} e^{-\ii k n}c^\dag_n\ket{\varnothing}=
\sum_{n=1}^N e^{-\ii k n}\tilde{c}^\dag_n\ket{\varnothing},
\\
\bra{k}_{\rm L}=&
\sum_{n=1}^N e^{+\frac12qn} e^{+\ii k n}c^{}_n\bra{\varnothing}=
\sum_{n=1}^N e^{+\ii k n}\tilde{c}^{}_n\bra{\varnothing},
\end{align}
\end{subequations}
up to normalization, and with energy eigenvalues
\begin{equation}\label{eq:energy}
E_k=2\cos{k},
\end{equation}
with momenta satisfying the boundary condition
$
e^{\frac12 \rho q N+\ii\phi}=
e^{\frac12 q N+\ii k N},
$
which gives
\begin{equation}\label{eq:momenta}
k_m=
\frac1N ({\phi}+ {2\pi m})
+ \frac\ii2 (1 - \rho)q
,
\end{equation}
for $m=0,\ldots,N-1\in\mathbb{Z}$.

Since the spectra are symmetric under $k\to-k$, it follows that for $\phi=0$ or $\pi$ (PBC or APBC for $\widetilde H$), the energy spectra for $\rho =\rho '$ and $\rho =2-\rho '$ coincide up to a relabeling of the energy level indexes $m\to -m \mod N$.
Indeed, choosing $\alpha_{\rm L}=1/\alpha_{\rm R}=\pm e^{\rho q N}$ is equivalent to choose $\alpha_{\rm L}=1/\alpha_{\rm R}=\pm e^{(2-\rho) q N}$ in \cref{eq:TransformedHamiltonian} up to exchanging the left and right boundary by inverting the lattice indexes $n\to N+1-n$ (space-inversion).
Hence, the Hamiltonian $\widetilde H$ in \cref{eq:TransformedHamiltonian} is unitary invariant up to the transformation
\begin{equation}\label{eq:symmetry}
\alpha_{\rm L}=1/\alpha_{\rm R}=e^{\ii\phi} e^{(2-\rho)qN}
\leftrightarrow
\alpha_{\rm L}=1/\alpha_{\rm R}=e^{\ii\phi} e^{\rho q N},
\end{equation}
for $\phi=0$ or $\pi$.
In particular, this mandates that the Hamiltonian $\widetilde H$ is Hermitian and symmetric up to space inversion symmetry $\mathcal{P}$ and time-reversal symmetry $\mathcal T$ for $\rho=1$ and $\phi=0,\pi$.
However, note that only the energy spectrum of the Hamiltonian $H$ is invariant up to this transformation, but not its eigenmodes, as one can immediately see from \cref{eq:eigenmodes}.

\Cref{fig:spectraeven} shows the trajectories of the eigenvalues of the Hamiltonians $H$ and $\widetilde H$ in the complex plane with $\alpha_{\rm L}$ and $\alpha_{\rm R}$ as in \cref{eq:alphas} as a function of $\rho$ in the interval $[0,2]$, given by \cref{eq:energy,eq:momenta} for different choices of $q$ and $\phi$ and with $N=10$.
Note the degeneracy of the spectra and the symmetry under \cref{eq:symmetry} for $\phi=0,\pi$.
Note that the condition in \cref{eq:specialBC} corresponds to a fine-tuning of the boundary hopping amplitudes, which becomes trivial in the continuum limit as $N\to\infty$, where $\alpha_{\rm L},\alpha_{\rm R}\to0$.
Hence, this effect is only physically relevant for finite lattices and can only be observed for small lattice sizes.
Additionally, the choice of $\phi$ changes the topology of the trajectories traced by the complex eigenvalues as a function of $\rho$.
These trajectories swirl around the origin for $\Im(q)\neq0$ with a rotation angle that increases with $\Im(q)$.

As discussed above, the case $\rho=1$ (see~\cref{eq:specialBC}) corresponds to real energy spectra.
Moreover, for all cases with $\rho>0$, the right and left eigenmodes of the Hamiltonian $H$ exhibit the skin effect, being exponentially localized on the left and on the right, respectively, for $\Re(q)>0$ (and vice versa for $\Re(q)<0$).
In the case $\rho=0$, the right and left eigenmodes become plane waves.
Conversely, for $\rho\neq1$, the right and left eigenmodes of the Hamiltonian $\widetilde H$ exhibit the skin effect, being exponentially localized on the opposite sides, becoming plane waves for $\rho=1$.
Note that for $0<\rho<1$, the right and left eigenmodes are localized respectively on the right and left boundary, while for $\rho>1$, they are localized on the opposite boundaries.
This reversal of the localization of the eigenmodes skin effect of $\widetilde H$ is realized by adjusting the hopping amplitudes between only two lattice sites.
\Cref{fig:skin} show the right and left eigenmodes of the Hamiltonians $H$ and $\widetilde H$.

Finally, for $\alpha_{\rm L}=\alpha_{\rm R}=0$, the right and left eigenmodes are in the form $\ket{k}_{\rm R}-\ket{-k}_{\rm R}$ and $\bra{k}_{\rm L}-\bra{-k}_{\rm L}$ with energies as in \cref{eq:energy} and with momenta satisfying the boundary condition $\sin(k (N+1))=0$, which gives
\begin{equation}
k_m=
\frac{\pi m}{N+1}
,
\end{equation}
as expected.


In conclusion, we have shown that generalized boundary conditions provide a means to control the spectral and localization properties of non-Hermitian lattices.
Indeed, tuning only one hopping amplitude allows one to induce exceptional points, drive transitions between real and complex spectra, and control the presence and direction of the non-Hermitian skin effect. 
Using exact similarity transformations, we identified conditions under which real spectra persist beyond open boundary conditions and demonstrated that both spectral and skin-effect properties can depend sensitively on system size. 
Our results highlight the strong boundary-dependence of non-Hermitian phenomena and open new routes for engineering finite-size non-Hermitian systems with tailored spectral and localization features.

\begin{acknowledgments}
P.~M. is partially supported by the Japan Society for the Promotion of Science (JSPS) Grant-in-Aid for Early-Career Scientists Grants No.~23K13028, Grant-in-Aid for Transformative Research Areas (A) KAKENHI Grant~No.~22H05111, and Grant-in-Aid for Transformative Research Areas (B) KAKENHI Grant~No.~24H00826.
\end{acknowledgments}


\end{document}